\documentclass[12pt]{article}
\usepackage[reqno]{amsmath}
\usepackage{bbm}
\usepackage{epsfig}
\usepackage{array}
\usepackage{float}
\usepackage{dsfont}
\usepackage{amstext}
\usepackage{amsfonts}

\usepackage{cite}

\usepackage{a4}

\usepackage{a4wide}
\def\be{\begin{equation}}
\def\ee{\end{equation}}
\def\gs{\mathrel{
   \rlap{\raise 0.511ex \hbox{$>$}}{\lower 0.511ex \hbox{$\sim$}}}}
\def\ls{\mathrel{
   \rlap{\raise 0.511ex \hbox{$<$}}{\lower 0.511ex \hbox{$\sim$}}}}

\newcommand{\ba}{\begin{array}{c}}
\newcommand{\baz}{\begin{array}{cc}}
\newcommand{\bad}{\begin{array}{ccc}}
\newcommand{\bav}{\begin{array}{cccc}}
\newcommand{\bea}{\begin{equation} \begin{array}{c}}
\newcommand{\eea}{ \end{array} \end{equation}}
\newcommand{\ea}{\end{array}}
\newcommand{\D}{\displaystyle}
\newcommand{\dms}{\mbox{$\Delta m^2_{\odot}$}}
\newcommand{\dma}{\mbox{$\Delta m^2_{\rm A}$}}

\newcommand{\MoreRep}[2]{\underline{\mbox{\textbf{#1}}} _{\mbox{\textbf{#2}}}}
\newcommand{\Groupname}[2]{$ {#1} _{#2} $}

\newcommand{\VEV}[1]{\langle #1 \rangle}
\begin{document}

\title{%\vspace{1cm}
\hfill {\small arXiv: 0903.0531 [hep-ph]} 
\vskip 1.cm
\bf %\large
Golden Ratio Prediction for Solar Neutrino Mixing\\[1cm]
}
\author{
Adisorn Adulpravitchai\thanks{email: 
\tt adisorn.adulpravitchai@mpi-hd.mpg.de}~~,~~
Alexander Blum\thanks{email: \tt alexander.blum@mpi-hd.mpg.de}
~~and~
Werner Rodejohann\thanks{email: \tt werner.rodejohann@mpi-hd.mpg.de} 
\\\\
{\normalsize \it Max--Planck--Institut f\"ur Kernphysik,}\\
{\normalsize \it  Postfach 103980, D--69029 Heidelberg, Germany}
}
\date{}
\maketitle
\thispagestyle{empty}
%\vspace{-0.8cm}
\begin{abstract}
\noindent 
It has recently been speculated that the solar neutrino 
mixing angle is connected to the golden 
ratio $\varphi$. Two such proposals have been made, 
$\cot \theta_{12} = \varphi$ and $\cos \theta_{12} = \varphi/2$. 
We compare these Ans\"atze and discuss a 
model leading to $\cos \theta_{12} = \varphi/2$ 
based on the dihedral group $D_{10}$. This symmetry is a natural
candidate because the angle in the expression 
$\cos \theta_{12} = \varphi/2$ is simply $\pi/5$, or 36 degrees.  
This is the exterior angle of a decagon and 
$D_{10}$ is its rotational symmetry group. 
We also estimate radiative corrections to 
the golden ratio predictions.

\end{abstract}

\newpage

\section{\label{sec:intro}Introduction}

The question what kind of flavor model underlies the 
peculiar features of lepton mixing is one of the dominating ones 
in contemporary theoretical neutrino physics. One hopes that 
precision measurements of the flavor parameters will provide 
hints towards the symmetry principle behind the apparent
regularities. We will in this paper discuss one intriguing example of
this line of thought.

All these issues are linked to the structure of the 
neutrino (and the charged lepton) mass matrix.  
Typically, the smallness of $|U_{e3}|$ and the close-to-maximality 
of $\theta_{23}$ are -- in the charged lepton basis -- 
attributed to the presence of an approximate 
$\mu$--$\tau$ symmetry:
\be
m_\nu = 
\left(
\bad 
A & B & B \\
\cdot & D & E \\
\cdot & \cdot & D 
\ea
\right) .
\ee
The eigenvector to the eigenvalue $D - E$ indeed is 
$(0,-1,1)^T$, but solar neutrino mixing is unconstrained by the 
matrix given above.  
If in addition to $\mu$--$\tau$ symmetry the 
condition $A + B = D + E$ holds, then the value 
$\sin^2 \theta_{12} = \frac 13$ is obtained: the infamous 
tri-bimaximal mixing \cite{tri}, which dominates the current 
theoretical literature on lepton flavor model building. 
However, comparing the tri-bimaximal 
mixing parameters ($\sin^2 \theta_{12} = \frac 13$, 
$\sin^2 \theta_{13} = 0$ and $\sin^2 \theta_{23} = \frac 12$) 
with the current best-fit, 1, 2 and $3\sigma$ ranges 
\cite{jose} (very similar results are found in \cite{bari}) 
\be
\bad 
\sin^2 \theta_{12} & \sin^2 \theta_{13} 
& \sin^2 \theta_{23} \\ \hline
0.304 & 0.01 & 0.50 \\
0.288 \div 0.326 & \le 0.026 & 0.44 \div 0.57 \\
0.27 \div 0.35 & \le 0.040 & 0.39 \div 0.63 \\
0.25 \div 0.37 & \le 0.056 & 0.36 \div 0.67 \\
\ea 
\ee
one notes that there is ample room for mixing scenarios 
%for $\theta_{12}$ 
other than tri-bimaximal mixing. 
In this short note we consider alternatives to tri-bimaximal mixing 
and focus on the fascinating possibility to link 
solar neutrino mixing with the 
golden ratio $\varphi = \varphi^2 - 1 
= \frac 12 \, (1 + \sqrt{5})$. 
Two such proposals have recently been made. The first one is 
\cite{other_golden,verrueckt,everett}
\be \label{eq:1}
\bav
\mbox{(A): } & \cot \theta_{12} = \varphi & \Rightarrow 
\sin^2 \theta_{12} = \frac{\D 1}{\D 1 + \varphi^2} 
= \frac{\D 2}{\D 5 + \sqrt{5}} & \simeq 0.276 \,.
\ea
\ee
The second possibility is \cite{WR}
\be \label{eq:2}
\bav 
\mbox{(B): } & \cos \theta_{12} = \frac{\D \varphi}{\D 2} & 
\Rightarrow \sin^2 \theta_{12} = 
\frac 14 \left(3 -  \varphi \right)
= \frac{\D 5 - \sqrt{5}}{\D 8}  
& \simeq 0.345 \,.
\ea
\ee
It can be seen that the both predictions 
lie within the current $2\sigma$ range\footnote{Actually, prediction 
(A) would lie very slightly outside the $2\sigma$ range of 
Ref.~\cite{bari}, which is $\sin^2 \theta_{12} = 0.278 \div 0.352$.}. 
The possibility (A) has first been noted in 
Ref.~\cite{other_golden}, and discussed in more detail in 
\cite{verrueckt}, where it was also mentioned that $A_5$ might 
be a candidate for the underlying flavor symmetry group. 
In this spirit a model 
based on $A_5$, which can lead to $\cot \theta_{12} = \varphi$, 
has been outlined in Ref.~\cite{everett}. The reason why 
$A_5$ is the candidate symmetry is because this group is isomorphic to the 
rotational group of the icosahedron and its geometrical features 
can be linked to the golden ratio. 
For instance, the 12 vertices of an icosahedron with edge-length 
2 have Cartesian coordinates $(0, \pm 1, \pm \varphi)$, 
$(\pm 1, \pm \varphi, 0)$ and $(\pm \varphi, 0, \pm 1)$.  
A peculiar feature of $\cot \theta_{12} = \varphi$ is that the angle 
gives also $\tan 2 \theta_{12} = 2$, and this 
can be obtained from a simple matrix proportional to 
\cite{verrueckt}
\be \label{eq:mnusimple}
m_\nu \propto  
\left(
\bad
0 & 1 \\
1 & 1 
\ea
\right)  . 
\ee
This matrix is invariant under a $Z_2$ symmetry generated by 
\cite{verrueckt}
\be \label{eq:Z2simple}
S = 
\frac{1}{\sqrt{5}}
\left(
\bad
-1 & 2 \\
2 & 1 
\ea
\right)  ,
\ee
where invariance is fulfilled when 
$S^T \, m_\nu \, S = m_\nu$.\\

Now consider the second golden ratio prediction $\cos \theta_{12} =
\varphi/2$, which corresponds simply to $\theta_{12} = \pi/5$. 
A mixing scenario based on this value 
was proposed with a purely phenomenological 
purpose in Ref.~\cite{WR}. 
A unified parametrization of both the CKM
and the PMNS matrix was constructed by choosing in addition to 
the lepton mixing angle $\theta_{12} = \pi/5$ a 
similar expression for the quark sector, namely 
$\theta_{12}^q = \pi/12$. 
The resulting value for the sine of the Cabibbo angle,  
$\sin \theta_{12}^q = (\sqrt{3}-1)/\sqrt{8}$, is also a 
simple algebraic and irrational number. The point made in
Ref.~\cite{WR} was that at zeroth order 
the CKM matrix is a 12-rotation with angle $\pi/12$, while 
the PMNS matrix is a 12-rotation with angle 
$\pi/5$ multiplied with an additional maximal (atmospheric) 23-rotation. 
To correct the 12-angles of the quark and lepton sectors 
to their respective best-fit values, one 
needs to multiply both zeroth order mixing matrices with 
a small 12-rotation. It turns out that one can achieve this 
with a universal (i.e., the same for quarks and
leptons) angle $\epsilon_{12} \simeq -0.03$ \cite{WR}. 

In the present letter we concentrate on the possible 
theoretical origin of the golden ratio prediction (B). 
We stress that flavor models based on the symmetry group $D_{10}$
are natural candidates to generate $\theta_{12} = \pi/5$. 
The dihedral group $D_{10}$ is the rotational symmetry group of a 
decagon and the exterior angle in a decagon is nothing but $\pi/5$, 
or 36 degrees. 
Indeed, we will present a model based on $D_{10}$ in the next Section
\ref{sec:A}. We remark that also $D_5$, the rotational symmetry group
of a regular pentagon, could be possible. In a pentagon the length of
a diagonal is $\varphi$ times the length of a side. The triangle
formed by the diagonal and two sides has one angle of $108^\circ$ (the
internal angle) and two angles with  $36^\circ$ each. 
However, here we focus on $D_{10}$ because it turns out 
that the vacuum alignment we need in our model 
is simplified due to the larger number of 
representations in $D_{10}$. Note that just as considering $A_5$ for 
the golden ratio prediction (A) was motivated by geometrical
considerations, the use of the (mathematically simpler) 
pentagon or decagon 
symmetry group is here motivated by prediction (B). These are examples for
the hope mentioned in the beginning, namely that precision
measurements may give us hints towards the underlying 
symmetry behind
flavor physics\footnote{Tri-bimaximal mixing is usually obtained with 
models based on $A_4$, the symmetry group of a tetrahedron \cite{A4}. 
%The vertices of a tetrahedron of edge-length $2\sqrt{2}$ 
%have Cartesian coordinates $(1, 1, 1)$, $(-1, -1, 1)$, $(-1, 1, -1)$, 
%$(1, -1, -1)$, which resembles the characteristic eigenvector 
%$(1,1,1)$ of $\nu_2$. More importantly, 
Here the angle between two faces 
(the dihedral angle) is $2 \theta_{\rm TBM}$, 
where $\sin^2 \theta_{\rm TBM} = \frac 13$.}.

%Note that it also holds that $\sin \, 666^\circ =
%-\varphi/2$, i.e., 
%solar neutrino mixing may be connected with 
%the {\tt Number of the Beast!!}  

The present paper is build up as follows: 
after discussing general symmetry properties of mass matrices 
with $\cos \theta_{12} = \varphi/2$ and an 
explicit $D_{10}$ model in Sec.~\ref{sec:A} 
we will in Section \ref{sec:RG} deal with renormalization 
group corrections to both golden ratio predictions 
(A) and (B), before we conclude in Section \ref{sec:concl}.

\section{\label{sec:A}Golden Ratio Prediction $\theta_{12} = \pi/5$ 
and Dihedral Groups}

We have seen in the Introduction that there is a simple $Z_2$ under 
which a mass matrix generating $\cot \theta_{12} = \varphi$ 
is invariant, see Eqs.~(\ref{eq:mnusimple}) and (\ref{eq:Z2simple}).  
The second golden ratio proposal (B) in Eq.~(\ref{eq:2}) 
corresponds to $\tan 2 \theta_{12} = 
\sqrt{1 + \varphi^2}/(\varphi - 1)$, and 
therefore it diagonalizes a less straightforward matrix. 
Nevertheless, in this case one can make use of $Z_2$ 
invariance as well, however the charged lepton sector has also 
to be taken into account. We will first discuss this for the simplified 
2-flavor case with symmetric mass matrices, 
before making the transition to dihedral groups 
and then to the explicit model based on $D_{10}$ that we will construct.\\
 
The generators of the $Z_2$ under which the neutrino 
mass matrix $m_\nu$ and the charged lepton mass matrix 
$m_\ell$ have to be invariant are 
\be \label{eq:Snuel}
S_{\nu, \ell} = 
\left(
\bad
0 & e^{-i \Phi_{\nu, \ell}} \\
e^{i \Phi_{\nu, \ell}} & 0 
\ea
\right)  ,
\ee
respectively. The matrices $m_\nu$ and $m_\ell$ 
are invariant when they have the following structure:  
\bea  \label{eq:structure}
m_{\nu, \ell} = 
\left(
\bad
A_{\nu, \ell} \, e^{i \Phi_{\nu, \ell}} & B_{\nu, \ell} \\
B_{\nu, \ell} & A_{\nu, \ell} \, e^{-i \Phi_{\nu, \ell}}
\ea
\right)  = 
\left(
\bad
e^{i \Phi_{\nu, \ell}} & 0 \\ 
0 & 1 
\ea
\right) 
\left(
\bad
A_{\nu, \ell} & B_{\nu, \ell} \\
B_{\nu, \ell} & A_{\nu, \ell} 
\ea
\right)
\left(
\bad
1 & 0 \\ 
0 & e^{-i \Phi_{\nu, \ell}} 
\ea
\right) \\
\equiv P_{\nu, \ell} 
\left(
\bad
A_{\nu, \ell} & B_{\nu, \ell} \\
B_{\nu, \ell} & A_{\nu, \ell} 
\ea
\right)
Q_{\nu, \ell} \, .
\eea
The inner matrix can be written as 
\be
\left(
\bad
A_{\nu, \ell} & B_{\nu, \ell} \\
B_{\nu, \ell} & A_{\nu, \ell} 
\ea
\right) = 
\tilde{U}_{\nu, \ell}^T \, 
{\rm diag }(A_{\nu, \ell} - B_{\nu, \ell}~,~
A_{\nu, \ell} + B_{\nu, \ell}) \, \tilde{U}_{\nu, \ell} \, ,
\mbox{ where } 
\tilde{U}_{\nu, \ell}
= 
\left( 
\bad
-\frac{1}{\sqrt{2}} & \frac{1}{\sqrt{2}} \\
\frac{1}{\sqrt{2}} & \frac{1}{\sqrt{2}}
\ea
\right) .
\ee
The total diagonalization matrices of $m_\nu$ and $m_\ell$ are 
$U_{\nu, \ell} = P_{\nu, \ell} \, \tilde{U}_{\nu, \ell}$ and 
the physical mixing matrix is their product 
$U = U_\ell^\dagger \, U_\nu = 
\tilde{U}_\ell^\dagger \, P_\ell^\dagger \, P_\nu \, \tilde{U}_\nu$.  
The 11-element is found to be 
\be \label{eq:mpi0}
|U_{e1}|^2 = \left|\cos \frac12 (\Phi_\nu - \Phi_\ell) \right|^2 \,.
\ee
The fact that a non-trivial phase matrix lies in between the two maximal
rotations $\tilde{U}_\ell^\dagger$ and $\tilde{U}_\nu$ is crucial. 
Obviously, at this stage any mixing angle can be generated.  
However, the observation made in Refs.~\cite{MPI} was that 
the phase factors in Eq.~(\ref{eq:Snuel}) 
can be linked to group theoretical flavor model
building with dihedral groups $D_n$. To make the connection from
Eq.~(\ref{eq:mpi0}) to dihedral groups, we note that 
the flavor symmetry $D_n$ has 2-dimensional representations 
$\MoreRep{2}{j}$, with ${\rm j} = 1, \ldots, \frac n2 - 1$ 
(${\rm j} = 1, \ldots, \frac{n-1}{2}$) for integer (odd) $n$, 
generated by 
\be
A = 
\left( 
\baz 
e^{2 \pi i \frac{\rm j}{n}} & 0 \\ 
0 & e^{-2 \pi i \frac{\rm j}{n}}
\ea
\right) \mbox{ and } 
B = 
\left( 
\baz 
0 & 1 \\ 
1 & 0 
\ea
\right). 
\ee 
$Z_2$ subgroups are generated by 
\be
B \, A^{k} = 
\left( 
\baz 
0 & e^{-2 \pi i  \frac{\rm j}{n} \, k} \\
e^{2 \pi i \frac{\rm j}{n} \, k} & 0 
\ea 
\right)
\ee
with integer $k$. This is just the required form of a 
$Z_2$ generator in Eq.~(\ref{eq:Snuel}).  
It is now possible to construct models in 
which the two fermions transform under the representation 
$\MoreRep{2}{j}$ of $D_n$, and $D_n$ 
is broken such that $m_\nu$ is left invariant under 
$B \, A^{k_\nu}$ and $m_\ell$ is left invariant under 
$B \, A^{k_\ell}$ \cite{MPI}. 
Consequently, the relation in Eq.~(\ref{eq:mpi0}) is obtained 
and we can identify 
\be \label{eq:ue1}
|U_{e1}|^2 = \left|\cos \pi 
\frac{\rm j}{n}
\left(k_\nu - k_\ell \right)
\right|^2 \, .
\ee
Hence, a natural candidate to implement  
the requested value of $\pi/5$ is e.g., $D_{10}$. 
This is no surprise given the observation that we made in the 
Introduction, namely that $\pi/5$ is the 
exterior angle of a decagon and that $D_{10}$ is its rotational
symmetry group.\\

We continue with an explicit model: 
we work in the framework of the MSSM without explicitly 
introducing right-handed neutrinos. Majorana masses for 
the light neutrinos are thus generated by an effective operator 
coupling to two Higgs vacuum expectation values (VEVs). 
We augment the MSSM by a flavor symmetry $D_{10} \times Z_5$. 
The symmetry $D_{10}$ is used for our prediction of the 
solar mixing angle, while the auxiliary Abelian symmetry $Z_5$
separates the charged lepton and neutrino sectors. 
Due to the flavor symmetry, no renormalizable Yukawa couplings 
are allowed for the charged leptons and the dimension 5 
operator giving mass to the neutrinos also vanishes. Mass 
for the leptons is generated by coupling them to gauge singlet 
flavons, which acquire VEVs and thereby break the flavor group. 
The charged lepton masses are thus generated by dimension 
5 operators, the neutrino masses by dimension 6 operators\footnote{ 
In a model including quarks, this may explain 
$m_{\tau} \ll m_t$ without invoking a large $\tan{\beta}$.}. 

The transformation properties of the MSSM leptons and Higgs 
fields, as well as the representations under which the 
flavons transform, are given in Table \ref{tab:particlesD10}. 
The multiplication table and the 
Clebsch-Gordan coefficients of $D_{10}$ are delegated to Appendix
\ref{app:d10}.
Note that the fermions and the flavons that couple to them 
are all in unfaithful representations of 
$D_{10}$ (i.e., in $\MoreRep{2}{2}$ and $\MoreRep{2}{4}$), 
so that here a $D_5$ structure would have 
sufficed. However, the full $D_{10}$ structure is 
needed to achieve the desired vacuum alignment.
We can continue by constructing the Yukawa superpotential, 
giving the leading order terms for both charged 
lepton and neutrino masses:
\begin{eqnarray} \label{eq:Y}
w_Y &=& y_1^e \, (l_1  \, e_2^c  \, \chi^e_2 + l_2  \, e_1^c  \, 
\chi^e_1)  \, \frac{h_d}{\Lambda} + y_2^e  \, (l_1  \, e_1^c  \, 
\rho^e_1 + l_2  \, e_2^c  \, \rho^e_2)  \, \frac{h_d}{\Lambda} 
+ y_3^e  \, (l_3  \, e^c_1  \, \chi^e_2 + l_3  \, e^c_2  \, \chi^e_1) 
 \, \frac{h_d}{\Lambda}\nonumber\\
&+& y_4^e  \, (l_1  \, e^c_3  \, \rho^e_2 + l_2  \, e^c_3  \, 
\rho^e_1)  \, \frac{h_d}{\Lambda} + y_5^e  \, l_3  \, e^c_3  \, 
\sigma^e  \, \frac{h_d}{\Lambda} \\
&+& y_1^{\nu}  \, l_1  \, l_2  \, \sigma^{\nu}  \, 
\frac{h_u^2}{\Lambda^2} + y_1^{\nu}  \, l_2  \, l_1  \, \sigma^{\nu} 
 \, \frac{h_u^2}{\Lambda^2} + y_2^{\nu}  \, (l_1  \, l_1  \, 
\chi^{\nu}_1 + l_2  \, l_2  \, \chi^{\nu}_2)  \, 
\frac{h_u^2}{\Lambda^2}+ y_3^{\nu}  \, l_3  \, l_3  \, 
\sigma^{\nu}  \, \frac{h_u^2}{\Lambda^2} \nonumber \, .
\end{eqnarray}

\begin{table}
\begin{center}
\begin{tabular}{|c||c|c|c|c||c||c|c|c|c||c|c|c|c|}\hline
Field & $l_{1,2} $ & $l_{3}$ & $ e^c_{1,2} $ & $e^c_3$ & $h_{u,d}$ & $\sigma^e$ & $\chi^e_{1,2}$ & $\xi^e_{1,2}$ & $\rho^e_{1,2}$ & $\sigma^{\nu}$ & $\varphi^{\nu}_{1,2}$ & $\chi^{\nu}_{1,2}$ & $\xi^{\nu}_{1,2}$ \\ 
\hline
\Groupname{D}{10} & $\MoreRep{2}{4}$ & $\MoreRep{1}{1}$ & $\MoreRep{2}{2}$ & 
$\MoreRep{1}{1}$ & $\MoreRep{1}{1}$  & $\MoreRep{1}{1}$ & 
$\MoreRep{2}{2}$ & $\MoreRep{2}{3}$ & $\MoreRep{2}{4}$ & $\MoreRep11$ & $\MoreRep21$ & $\MoreRep22$ & $\MoreRep23$ \\
\Groupname{Z}{5} & $\omega$ & $\omega$ & $\omega^2$ & $\omega^2$ & $1$ & $\omega^2$ & $\omega^2$ & $\omega^2$
& $\omega^2$ & $\omega^3$ & $\omega^3$ & $\omega^3$ & $\omega^3$\\
\hline
\end{tabular}
\normalsize
%\begin{minipage}[t]{15cm}
\caption{Particle content of the 
$D_{10}$ model: 
$l_i$ are the three left-handed lepton doublets, 
$e_i^c$ are the right-handed charged lepton singlets and  
$h_{u,d}$ are the MSSM Higgs doublets. Furthermore we have flavons 
$\sigma^e$, $\chi^e_{1,2}$, $\xi^e_{1,2}$, $\rho^e_{1,2}$, 
$\sigma^{\nu}$, $\varphi^{\nu}_{1,2}$, $\chi^{\nu}_{1,2}$ 
and $\xi^{\nu}_{1,2}$ which only transform under 
$D_{10} \times Z_5$. The phase $\omega = e^{\frac{2 \pi i}{5}}$ is 
the fifth root of unity. 
\label{tab:particlesD10}}
%\end{minipage}
\end{center}
\end{table}

\noindent As we will show below in Appendix 
\ref{sec:vev}, introducing appropriate ``driving fields'' and 
minimizing the flavon superpotential leads to the following 
VEVs for the flavons:
\begin{equation}
\label{eq:VEVs_e}
\left( \begin{array}{c} \VEV{\chi_1^e} \\ \VEV{\chi_2^e} \end{array} \right) = v_e \left( \begin{array}{c} 1 \\ e^{\frac{2 \pi i k}{5}} \end{array} \right)  ,~
\left( \begin{array}{c} \VEV{\xi_1^e} \\ \VEV{\xi_2^e} \end{array} \right) = w_e \left( \begin{array}{c} 1 \\ e^{\frac{3 \pi i k}{5}} \end{array}\right) ,~
\left( \begin{array}{c} \VEV{\rho_1^e} \\ \VEV{\rho_2^e} \end{array}
\right) = z_e \left( \begin{array}{c} 1 \\ e^{\frac{4 \pi i k}{5}}
\end{array} \right) ,
\end{equation}
where $k$ is an odd integer between 1 and 9, and 
\begin{equation}
\label{eq:VEVs_nu}
\left( \begin{array}{c} \VEV{\varphi_1^{\nu}} \\ \VEV{\varphi_2^{\nu}} \end{array} \right) = v_{\nu} \left( \begin{array}{c} 1 \\ 1 \end{array} \right) \; , \;\; 
\left( \begin{array}{c} \VEV{\chi_1^{\nu}} \\ \VEV{\chi_2^{\nu}} \end{array} \right) = w_{\nu} \left( \begin{array}{c} 1 \\ 1 \end{array}\right) \; , \;\;
\left( \begin{array}{c} \VEV{\xi_1^{\nu}} \\ \VEV{\xi_2^{\nu}}
\end{array} \right) = z_{\nu} \left( \begin{array}{c} 1 \\ 1
\end{array} \right) .
\end{equation}
The VEVs of the singlet flavons $\langle \sigma^e \rangle = x_e$
and $\langle \sigma^{\nu} \rangle = x_{\nu}$ are assumed to be 
also non-vanishing. The VEV structure leads to the 
following mass matrices:
\bea \label{eq:mnumlep}
m_\ell 
= \frac{\D \langle h_d \rangle}{\D \Lambda} \, 
\left( 
\begin{array}{ccc}
y_2^e \, z_e & y_1^e  \, v_e  \, 
e^{\frac{2 \pi i k}{5}} & y_4^e  \, z_e  \, 
e^{\frac{4 \pi i k}{5}} \\ 
y_1^e  \, v_e  & y_2^e  \, z_e  \, 
e^{\frac{4 \pi i k}{5}} &  y_4^e  \, z_e \\ 
y_3^e  \, v_e  \, e^{\frac{2 \pi i k}{5}} &  y_3^e  \, v_e & y_5^e  
\, x_e 
\end{array} 
\right) , \\ 
m_{\nu} = \frac{\D \langle h_u \rangle^2}{\D \Lambda^2} \, 
\left( 
\begin{array}{ccc}  
y_2  \, w_{\nu} & y_1  \, x_{\nu} & 0 \\ 
\cdot & y_2  \, w_{\nu} &  0 \\ 
\cdot &  \cdot & y_3  \, x_{\nu} 
\end{array} 
\right) .
\eea
To see that
indeed the golden ratio prediction is obtained from the above two
matrices, note that for the choice 
$k = 3$ the relevant matrix $m_\ell \, m_\ell^\dagger$ 
takes the form 
\be \label{eq:mlml}
m_\ell \, m_\ell^\dagger = 
\left( 
\bad 
A & B \, e^{-2 i \Phi} & D \, e^{i (\delta - \Phi)} \\ 
B \, e^{2 i \Phi} & A & D \, e^{i (\delta + \Phi)} \\
D \, e^{-i (\delta - \Phi)} & D \, e^{-i (\delta + \Phi)} & G 
\ea
\right) ,\mbox{ where } \Phi = \frac{4 \pi}{5} \,.
\ee
The quantities $A, B, D, G$ are real and positive, 
$\delta$ is a phase. 
To obtain the golden ratio prediction for the solar 
mixing angle, we have to set 
in Eqs.~(\ref{eq:VEVs_e},\ref{eq:mnumlep}) 
$k = 3$ or $k = 7$. From the other 
possibilities $k = 1$ or $k = 9$ would give a solar 
mixing angle of $\frac{2 \pi}{5}$, while $k = 5$ would give a 
vanishing solar mixing angle. This small number of 
degeneracies can not be resolved by the flavon potential. 
Looking at the last matrix $m_\ell \, m_\ell^\dagger$ 
in Eq.~(\ref{eq:mlml}), one immediately recognizes the 
$Z_2$-invariance of the 
upper left 12-block, which is just the invariance we 
were seeking for, see Eq.~(\ref{eq:structure}). To be precise, 
the $D_{10}$ was broken in a way that $m_\ell \, m_\ell^\dagger$ 
is left invariant under $B \, A^3$, while  
the neutrino mass 
matrix $m_\nu$ is left invariant under $B \, A^0 = B$. 
Inserting this in Eq.~(\ref{eq:ue1}), where we have to 
set ${\rm j} = 4$ because 
the first and second left-handed lepton doublets 
transform as $\MoreRep{2}{4}$, we expect 
$|U_{e1}|^2 = |\cos \frac{6}{5} \pi|^2 $, which is indeed equivalent to 
an angle of $\pi/5$. We will explicitly check this in the following. 
Diagonalizing $m_\ell \, m_\ell^\dagger$  
with the relation $U_\ell^\dagger \, m_\ell \, m_\ell^\dagger \, 
U_\ell = {\rm diag}(m_e^2 ,\, m_\mu^2, \, m_\tau^2)$ 
is achieved with the matrix 
\be
U_\ell = {\rm diag}(e^{-2 i \Phi}, 1,e^{-i (\Phi + \delta)} ) \, 
\left( \bad 
-\sqrt{\frac 12} & \sqrt{\frac 12} & 0\\
\sqrt{\frac 12} & \sqrt{\frac 12} & 0 \\ 
0 & 0 & 1 
\ea
\right) 
\left( \bad 
1 & 0 & 0 \\
0 & \cos \theta_{23} & \sin \theta_{23} \\
0 & - \sin \theta_{23} & \cos \theta_{23}
\ea
\right) .
\ee
The diagonal phase matrix on the left is crucial. 
The rotation angle in the 23-axis is given by 
\be \label{eq:theta23} 
\tan 2 \theta_{23} =\frac{ 2\sqrt{2} \, D}{G - A - B} \, , 
\ee
and the charged lepton masses are given by 
\bea\label{eq:mleps}
m_e^2 = A - B ~,~~
m_{\mu, \tau}^2 = \frac 12\left[ (A + B + G) \pm w \, 
(A + B - G) \right] \, ,\\[0.2cm] \mbox{ where } 
w = \sqrt{1 + 8 \, D^2 /(A + B - G)^2} \, .
\eea
The neutrino mass matrix is diagonalized via 
$U_\nu^\dagger \, m_\nu \, U_\nu^\ast = m_\nu^{\rm diag}$,  
with 
\be
U_\nu = 
\left(
\bad 
-\sqrt{\frac 12} & \sqrt{\frac 12} & 0\\
\sqrt{\frac 12} & \sqrt{\frac 12} & 0 \\ 
0 & 0 & 1 
\ea
\right) P \, .
\ee
The eigenvalues have in general non-trivial phases which are taken into 
account in the diagonal matrix $P$, and their absolute values are 
\be \label{eq:mnus}
m_1 = \frac{\D \langle h_u \rangle^2}{\D \Lambda^2} 
|y_2 \, w_\nu - y_1 \, x_\nu|~,~~ 
m_2 = \frac{\D \langle h_u \rangle^2}{\D \Lambda^2} 
|y_2 \, w_\nu + y_1 \, x_\nu|~,~~ 
m_3 = \frac{\D \langle h_u \rangle^2}{\D \Lambda^2} 
|y_3 \, x_\nu| \, . 
\ee
We note that the 
model makes no predictions about the neutrino masses or their ordering. 
Nevertheless, one can easily convince oneself that the number of free
parameters in the model is
enough to fit the neutrino and charged lepton masses, as well as
the large atmospheric neutrino mixing angle $\theta_{23}$. The 
model does in general not predict $\theta_{23}$ to be maximal, 
which is not an issue given the fact that it is the lepton 
mixing parameter with the largest allowed range. However, 
maximal mixing is 
compatible with the model. We have $\theta_{23} = \pi/4$ when 
$G = A + B$, in which case $m_{\mu, \tau}^2 = 
A + B \mp \sqrt{2} \, D$ and $m_e^2$ 
as in Eq.~(\ref{eq:mleps}). The fact that there is not more 
predictivity can be traced to the fact that there is a comparably
large number of flavon fields required in order to make the model
work. This is the price one unfortunately has to pay if one insists 
in the rather peculiar value of $\theta_{12}$. Given the fact that current 
data allows for this very interesting possibility, one should 
nevertheless pursue the task of constructing models leading
to it. 

The final PMNS matrix is 
\be
U = U_\ell^\dagger \, U_\nu \, .
\ee
One finds that $U_{e3}$ is vanishing and that 
atmospheric neutrino mixing is governed by 
$\tan 2 \theta_{23}$ given by Eq.~(\ref{eq:theta23}).  
As mentioned above, the PMNS matrix has a non-trivial phase matrix 
including $\Phi$ in between the two maximal 12-rotations, one of which
stems from $U_\ell$, the other from $U_\nu$. 
As discussed above, this is the origin of the required result. 
Indeed, the $12$-element of $U$ is 
\be
|U_{e2}|^2 = \sin^2 \Phi = \sin^2 \pi/5 \, ,
\ee
and due to $U_{e3}=0$ this is just $\sin^2 \theta_{12}$. We have thus 
achieved our goal of predicting $\theta_{12} = \pi/5$. 
As discussed in Appendix \ref{sec:vev}, higher order 
corrections to the scenario, as well as flavor 
changing neutral currents, can be estimated to give only very small 
contributions.

\section{\label{sec:RG}Renormalization Corrections to the 
Golden Ratio Predictions}

It is worth discussing renormalization group (RG) effects to the golden
ratio predictions, because any 
symmetry leading to the predictions discussed in this paper could 
presumably be operating at a high energy scale $\Lambda$, and the
observables have to be evolved down to the low energy scale
$\lambda$. Note that RG corrections to 
$|U_{e3}|$ and $\theta_{23}$ are typically
suppressed with respect to the running of $\theta_{12}$ by a factor of
$\dms/\dma$. As the initial values of both $|U_{e3}|$ and $\theta_{23}$
need not to be specified here (other than being small or close to
maximal, respectively) we do not comment on their RG-shift. 
We will stay here model-independent and estimate the 
corrections as a function of the unknown neutrino mass values and
ordering. 
An expression for $\dot \theta_{12}$, where the dot denotes the 
derivative with respect to $t = \ln \mu/\mu_0$ with 
$\mu$ the renormalization scale, is given e.g., 
in \cite{RG}. 
One can therefrom estimate the shift for the solar 
neutrino mixing angle: 
\be \label{eq:dth12}
\theta_{12} \simeq \theta_{12}^0 + k_{12} \, \epsilon_{\rm RG}\,,
\ee 
where $\theta_{12}^0$ is the initial value of $\theta_{12}$ 
(here given by Eq.~(\ref{eq:1}) or (\ref{eq:2})) and 
\be
\epsilon_{\rm RG} \equiv c \,  
\frac{m_\tau^2}{16 \pi^2 \, v_u^2} \, \ln \frac \Lambda\lambda
\ee
with $v_u = 246$ GeV, $c = -\frac 32$ 
in the SM and $(1 + \tan^2 \beta)$ in the MSSM. 
Neutrino physics is included in 
\be \label{eq:k12}
k_{12} = \sin 2 \theta_{12}^0 \, \sin^2 \theta_{23}^0 \, 
\frac{\left|m_1 + m_2 \, e^{2i\alpha}\right|^2}{\dms}\,.
\ee 
Consequently, from Eq.~(\ref{eq:dth12}) one 
finds\footnote{Inserting 
$\sin^2 \theta_{12}^0 = \frac 13$ and 
$\sin^2 \theta_{23}^0 = \frac 12$ in the following 
and the last expression reproduces the results from Ref.~\cite{DGR}.}
\be \label{eq:RG}
\sin^2 \theta_{12} \simeq \sin^2 \theta_{12}^0 + 
k_{12} \, \epsilon_{\rm RG} \, \sin 2 \theta_{12}^0 \,.
\ee
Note that the Majorana phase $\alpha$ can suppress the running. 
As well known, $\theta_{12}$ decreases in the SM and 
increases in the MSSM, independent on the sign of 
$\dma$. 
The following numerical estimates are done 
with $\sin^2 \theta_{23}^0 = \frac 12$, $\Lambda/\lambda = 10^{10}$
and with $\dms, \dma$ fixed for simplicity 
at their current best-fit values \cite{jose}. 
In the normal hierarchy (NH, $m_3 \simeq \sqrt{\dma}$, 
$m_2 \simeq \sqrt{\dms} \gg m_1$) the running in the SM is
completely negligible. In case of the MSSM, even for 
$\tan \beta = 40$ the shift in $\sin^2 \theta_{12}$ is not more 
than 1.5 \%. This changes in the inverted hierarchy 
(IH, $m_2 \simeq m_1 \simeq \sqrt{\dma} \gg m_3$), where in the MSSM 
and $\tan \beta = 10$ the value of $\sin^2 \theta_{12}$ can 
increase by around 10 \%. In the SM, again, the shift is with 
less than half a percent not measurable. For quasi-degenerate
neutrinos with a common mass scale of 0.2 eV the SM allows 
shifts of around 3 \%, whereas in the MSSM the shift can 
be as large as the value of $\sin^2 \theta_{12}$, even for small
values of $\tan \beta = 5$. 
\begin{figure}[ht]%\vspace{-1cm}
\begin{center}
\epsfig{file=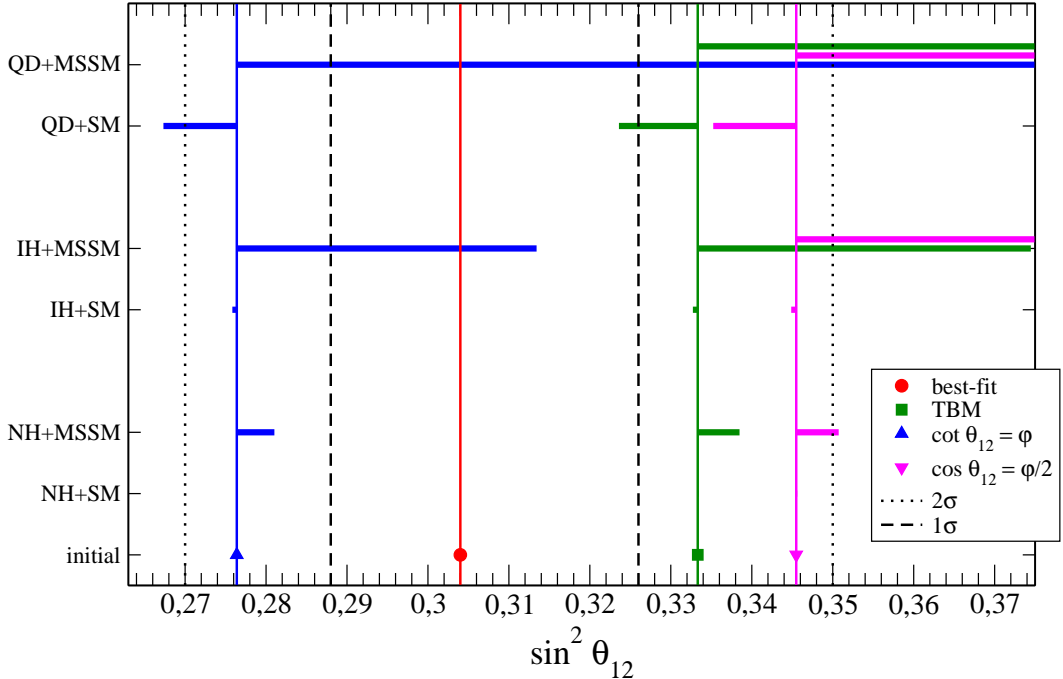,width=14cm,height=9cm} 
\end{center}
\caption{\label{fig:gold}RG-induced 
shifts on $\sin^2 \theta_{12}$, estimated from  
Eq.~(\ref{eq:RG}), for the two golden 
ratio proposals and for tri-bimaximal mixing (TBM). The current 
best-fit value, as well as the 1 and $2\sigma$ ranges are 
also indicated. 
In case of the MSSM we have taken $\tan \beta = 40$ 
for a normal hierarchy (NH), $\tan \beta = 10$ 
for the inverted hierarchy (IH) and $\tan \beta = 5$ 
for quasi-degenerate neutrinos (QD) with a 
mass scale of 0.2 eV. The line for 
the SM and a normal hierarchy cannot be seen because the effect is
too small. } 
\end{figure}
We illustrate this in 
Fig.~\ref{fig:gold}, where we used Eq.~(\ref{eq:RG}) to  
show the RG-induced shifts of $\sin^2 \theta_{12}$ 
for a normal mass hierarchy (SM and 
MSSM with $\tan \beta = 40$), an inverted hierarchy 
(SM and MSSM with $\tan \beta = 10$), as well as 
quasi-degenerate neutrinos (QD, smallest mass 0.2 eV for the 
SM and MSSM with $\tan \beta = 5$). In case of a normal and inverted
hierarchy we have chosen (at high scale) 0.001 eV 
for the smallest neutrino mass. To a
good approximation and unless in the MSSM $\tan \beta$ is very large, 
the running of the neutrino masses can be 
described by a rescaling, with basically no dependence on the other 
neutrino parameters \cite{RG}. Because $k_{12}$ from 
Eq.~(\ref{eq:k12}) has the masses appearing in the denominator 
and numerator, their running cancels in our approximation as long as 
$|k_{12} \, \epsilon_{\rm RG}| \ll 1$. 
The range of the corrections in 
Fig.~\ref{fig:gold} is due to the unknown Majorana phases. 
For illustration, we also 
include the shifts for tri-bimaximal mixing. 

To bring $\theta_{12}$ very close to the best-fit value, the prediction 
(A) requires the MSSM and IH or QD, while prediction (B) 
(and tri-bimaximal mixing) requires the SM with rather large 
neutrino masses. 
If future data leads to more precise determinations of $\sin^2
\theta_{12}$ and other neutrino parameters,  
one will be able to rule out some of the existing possibilities.

\section{\label{sec:concl}Summary}

Precision flavor data may give hints towards the underlying physics. 
We have stressed in this paper 
that current data implies that the golden ratio 
$\varphi$ can be connected to solar neutrino mixing. 
With $\cot \theta_{12} = \varphi$ and 
$\cos \theta_{12} = \varphi/2$ there are two appealing
possibilities, not too far away from current best-fit values 
and compatible
with current $2\sigma$ ranges. 
We have compared these values, estimated radiative 
corrections and in particular proposed a model based 
on the dihedral group 
$D_{10}$ leading to the relation $\cos \theta_{12} = \varphi/2$. 
%We have taken advantage of the fact that $D_{10}$ can be broken 
%down to non-trivial $Z_2$ subgroups. 
The angle leading to $\cos \theta_{12} = \varphi/2$ is  
$\theta_{12} = \pi/5$ and closely linked to the symmetry of a 
decagon, which naturally leads one to consider its 
rotational symmetry group $D_{10}$.

%\newpage
\vspace{0.3cm}
\begin{center}
{\bf Acknowledgments}
\end{center}
This work was supported by the ERC under the Starting Grant 
MANITOP and by the Deutsche Forschungsgemeinschaft 
in the Transregio 27 ``Neutrinos and beyond -- weakly interacting 
particles in physics, astrophysics and cosmology'' (W.R.). 
A.B.~acknowledges support from the Studienstiftung des 
Deutschen Volkes.

\renewcommand{\theequation}{B\arabic{equation}}
\setcounter{equation}{0}
\renewcommand{\thetable}{A\arabic{table}}
\setcounter{table}{0}

\begin{appendix}

%%%%%%%%%%%%%%%%%%%%%%%%%%%%%%%%%%%%%%%%%%%%%%

\section{Multiplication Rules and Clebsch-Gordan Coefficients of $D_{10}$}
\label{app:d10}

%%%%%%%%%%%%%%%%%%%%%%%%%%%%%%%%%%%%%%%%%%%%%%

\begin{table}
\begin{center}
\begin{tabular}{|c|cccc|}
\hline
$\times$&$\MoreRep{1}{1}$&$\MoreRep{1}{2}$&$\MoreRep{1}{3}$&$\MoreRep{1}{4}$\\
\hline
$\MoreRep{1}{1}$        &$\MoreRep{1}{1}$       &$\MoreRep{1}{2}$       &$\MoreRep{1}{3}$       &$\MoreRep{1}{4}$\\
$\MoreRep{1}{2}$        &$\MoreRep{1}{2}$       &$\MoreRep{1}{1}$       &$\MoreRep{1}{4}$       &$\MoreRep{1}{3}$\\
$\MoreRep{1}{3}$        &$\MoreRep{1}{3}$       &$\MoreRep{1}{4}$       &$\MoreRep{1}{1}$       &$\MoreRep{1}{2}$\\
$\MoreRep{1}{4}$        &$\MoreRep{1}{4}$       &$\MoreRep{1}{3}$       &$\MoreRep{1}{2}$       &$\MoreRep{1}{1}$\\
\hline
\end{tabular}\\
\vspace{0.3cm}
\begin{tabular}{|c|cccc|}
\hline
$\times$&$\MoreRep{2}{1}$&$\MoreRep{2}{2}$&$\MoreRep{2}{3}$&$\MoreRep{2}{4}$\\
\hline
$\MoreRep{2}{1}$        & $\MoreRep{1}{1}+\MoreRep12+\MoreRep22$       &$\MoreRep{2}{1}+\MoreRep23$       &$\MoreRep{2}{2}+\MoreRep24$       &$\MoreRep{1}{3}+\MoreRep14+\MoreRep23$\\
$\MoreRep{2}{2}$        &$\MoreRep{2}{1}+\MoreRep23$        &$\MoreRep{1}{1}+\MoreRep12+\MoreRep24$       &$\MoreRep{1}{3}+\MoreRep14+\MoreRep21$       &$\MoreRep{2}{2}+\MoreRep24$\\
$\MoreRep{2}{3}$        &$\MoreRep{2}{2}+\MoreRep24$        &$\MoreRep{1}{3}+\MoreRep14+\MoreRep21$       &$\MoreRep{1}{1}+\MoreRep12+\MoreRep24$       &$\MoreRep{2}{1}+\MoreRep23$\\
$\MoreRep{2}{4}$        & $\MoreRep{1}{3}+\MoreRep14+\MoreRep23$       &$\MoreRep{2}{2}+\MoreRep24$       &$\MoreRep{2}{1}+\MoreRep23$       &$\MoreRep{1}{1}+\MoreRep12+\MoreRep22$\\
\hline
\end{tabular}
\begin{equation*}
\MoreRep1{1,2} \times \MoreRep2j = \MoreRep2j \; , \; \; \MoreRep1{3,4} \times \MoreRep2j = \MoreRep2{5-j}
\end{equation*}
\caption{Multiplication rules for the dihedral group $D_{10}$, 
which has four two-dimensional and four 1-dimensional irreducible
representations. 
\label{tab:d10}}
\end{center}
\end{table}

\noindent We present here the Clebsch-Gordan coefficients for 
$D_{10}$. The multiplication rules for the Kronecker products are 
given in Table \ref{tab:d10}. 
For $s_i \sim \MoreRep{1}{i}$ and $(a_1,a_2)^{T} \sim \MoreRep{2}{j}$ we find

\begin{equation}\nonumber
\left( \begin{array}{c} s_1 a_1 \\ s_1 a_2
\end{array} \right) \sim \MoreRep{2}{j} \;\; , \;\;\;
\left( \begin{array}{c} s_2 a_1 \\ -s_2 a_2
\end{array} \right) \sim \MoreRep{2}{j} \;\; , \;\;\;
\left( \begin{array}{c} s_3 a_2 \\ s_3 a_1
\end{array} \right) \sim \MoreRep{2}{5-j} \;\;\; \mbox{and} \;\;\;
\left( \begin{array}{c} s_4 a_2 \\ -s_4 a_1
\end{array} \right) \sim \MoreRep{2}{5-j} \;\; .
\end{equation}
\noindent The Clebsch-Gordan coefficients for the 
product of $(a_{1},a_{2})^{T}$ with $(b_{1},b_{2})^{T}$, both in  
$\sim \MoreRep{2}{i}$, read
\begin{eqnarray}\nonumber
&& a_{1} b_{2} + a_{2} b_{1} \sim \MoreRep{1}{1} \; , \;\;
 a_{1} b_{2} - a_{2} b_{1} \sim \MoreRep{1}{2} \; , 
\\ \nonumber
&& \left( \begin{array}{c}
        a_{1} b_{1}\\
        a_{2} b_{2}
        \end{array}
        \right) \sim \MoreRep{2}{j} \;\;\; \mbox{or} \;\;\;
 \left( \begin{array}{c}
        a_{2} b_{2}\\
        a_{1} b_{1}
        \end{array}
        \right) \sim \MoreRep{2}{j} \, ,
\end{eqnarray}
\noindent depending on whether $\rm i =1,2$
or $\rm i = 3,4$.
For the two doublets $(a_{1}, a_{2})^{T} \sim \MoreRep{2}{i}$ and
$(b_{1}, b_{2})^{T} \sim \MoreRep{2}{j}$ we find for 
$\rm i + j \neq 5$

\begin{eqnarray}\nonumber
&& \left( \begin{array}{c}
        a_{1} b_{2}\\
        a_{2} b_{1}
        \end{array}
        \right) \sim \MoreRep{2}{k} \;\;\; (\rm k=i-j) \;\;\; \mbox{or} \;\;\;
 \left( \begin{array}{c}
        a_{2} b_{1}\\
        a_{1} b_{2}
        \end{array}
        \right) \sim \MoreRep{2}{k} \;\;\; (\rm k=j-i) \, ,
\\ \nonumber
&& \left( \begin{array}{c}
        a_{1} b_{1}\\
        a_{2} b_{2}
        \end{array}
        \right) \sim \MoreRep{2}{l} \;\;\;  (\rm l=i+j) \;\;\;\;\; \mbox{or} \;\;\;
 \left( \begin{array}{c}
        a_{2} b_{2}\\
        a_{1} b_{1}
        \end{array}
        \right) \sim \MoreRep{2}{l} \;\;\; (\rm l=10 -(i+j)) \, .
\end{eqnarray}
\noindent If $\rm i+j=5$ holds the covariants read
\begin{eqnarray}\nonumber
&& a_{1} b_{1} + a_{2} b_{2} \sim \MoreRep{1}{3} \; , \;\;
 a_{1} b_{1} - a_{2} b_{2} \sim \MoreRep{1}{4} \; , 
\\ \nonumber
&& \left( \begin{array}{c}
        a_{1} b_{2}\\
        a_{2} b_{1}
        \end{array}
        \right) \sim \MoreRep{2}{k} \;\;\; \mbox{or} \;\;\;
 \left( \begin{array}{c}
        a_{2} b_{1}\\
        a_{1} b_{2}
        \end{array}
        \right) \sim \MoreRep{2}{k} \; .
\end{eqnarray}
\noindent Again, the first case is relevant for $\rm k=i-j$, while the 
second one is valid for
$\rm k=j-i$.

\section{VEV Alignment of the $D_{10} \times Z_5$ Model}
\label{sec:vev}

\renewcommand{\thetable}{B\arabic{table}}
\setcounter{table}{0}

\noindent To obtain the necessary vacuum alignment in the 
flavon potential, we need to introduce a $U(1)_R$ and driving 
fields \cite{drivingfields}. Regular $R$-parity is a subgroup of the $U(1)_R$. 
To ensure a supersymmetric Lagrangian, the superpotential 
must have a $U(1)_R$ charge of 2. The superfields 
containing the SM fermions have an $R$-charge of 1, 
while the Higgs fields have an $R$-charge of zero. 
Hence, for the Yukawa superpotential given in Eq.~(\ref{eq:Y}) 
to be viable, the flavons also need to have a vanishing $R$-charge. 
Consequently, for the flavon superpotential one needs to introduce 
additional flavor-charged fields, having an $R$-charge of 2. 
The transformation properties of these driving fields 
are given in Table \ref{tab:driving}. 
\begin{table}
\begin{center}
\begin{tabular}{|c||c|c|c||c|c|c|}\hline
Field & $\psi^{0e}$ & $\varphi^{0e}_{1,2}$ & $\xi^{0e}_{1,2}$ & $\psi^{0\nu}$ & $\chi^{0\nu}_{1,2}$ & $\xi^{0\nu}_{1,2}$\\
\hline
\Groupname{D}{10} & $\MoreRep{1}{3}$ & $\MoreRep{2}{1}$ & $\MoreRep{2}{3}$ & 
$\MoreRep{1}{4}$ & $\MoreRep{2}{2}$  & $\MoreRep{2}{3}$ \\
\Groupname{Z}{5} & $\omega$ & $\omega$ & $\omega$ & $\omega^4$ & $\omega^4$ & $\omega^4$ \\
\hline
\end{tabular}
\normalsize
\begin{minipage}[t]{15cm}
\caption{Transformation properties of the driving fields 
under $D_{10} \times Z_5$. Again $\omega$ is the 
fifth root of unity $e^{\frac{2 \pi i}{5}}$. 
\label{tab:driving}}
\end{minipage}
\end{center}
\end{table}
The flavon superpotential can then be divided into two parts
\begin{equation}
w_f = w_{f,e} + w_{f, \nu} \, ,
\end{equation}
where $w_{f,e}$ and $w_{f,\nu}$ are responsible for the vacuum 
alignment of the flavons contributing to the charged 
lepton and neutrino masses, respectively. 
We begin by considering the charged lepton part: 
\begin{eqnarray}
w_{f,e} &=& a_e  \, (\chi^e_1  \, \xi^e_1 + \chi^e_2  \, 
\xi^e_2)  \, \psi^{0e} + b_e  \, (\chi^e_1  \, \xi^e_2  \, 
\varphi^{0e}_1 + \chi^e_2  \, \xi^e_1  \, \varphi^{0e}_2) 
+ c_e  \, (\xi^e_1  \, \rho^e_2  \, 
\varphi^{0e}_1 + \xi^e_2  \, \rho^e_1  \, \varphi^{0e}_2) 
\nonumber\\
&+& d_e  \, (\xi^e_2  \, \xi^{0e}_1 + \xi^e_1  \, \xi^{0e}_2)  \, 
\sigma^e + f_e  \, (\xi^e_1  \, \rho^e_1  \, \xi^{0e}_1 
+ \xi^e_2  \, \rho^e_2  \, \xi^{0e}_2) \, .
\end{eqnarray}
As the flavor symmetry is broken at a high scale, the scalar 
potential can be minimized in the supersymmetric limit. The 
flavons and driving fields are not charged under any gauge group, 
so the scalar potential is given by the F-terms alone. Hence, we can 
determine the supersymmetric minimum of the potential by 
setting the F-terms of the driving fields to zero: 
\begin{eqnarray}\nonumber
\frac{\partial w_{f,e}}{\partial \psi^{0e}} &=& a_e  \, 
(\chi^e_1  \, \xi^e_1 + \chi^e_2  \, \xi^e_2) = 0 \, ,
\\ \nonumber
\frac{\partial w_{f,e}}{\partial \varphi_1^{0e}} &=& b_e  \, 
\chi^e_1  \, \xi^e_2   + c_e  \, \xi^e_1  \, \rho^e_2  = 0 \, ,
\\ \nonumber
\frac{\partial w_{f,e}}{\partial \varphi_2^{0e}} &=& b_e  \, 
\chi^e_2  \, \xi^e_1  + c_e  \, \xi^e_2  \, \rho^e_1   = 0 \, ,
\\ \nonumber
\frac{\partial w_{f,e}}{\partial \xi_1^{0e}} &=& d_e  \, \xi^e_2 
\sigma^e + f_e  \, \xi^e_1  \, \rho^e_1  = 0 \, ,
\\ \nonumber
\frac{\partial w_{f,e}}{\partial \xi_2^{0e}} &=& d_e  \, 
\xi^e_1  \, \sigma^e + f_e  \, \xi^e_2  \, \rho^e_2  = 0 \, .
\end{eqnarray}
Similarly, from the neutrino part 
\begin{eqnarray}
w_{f,\nu} &=& a_{\nu}  \, (\chi^{\nu}_1  \, \xi^{\nu}_1 - 
\chi^{\nu}_2  \, \xi^{\nu}_2)  \, \psi^{0\nu} + b_{\nu}  \, 
(\varphi^{\nu}_1  \, \chi^{\nu}_1  \, \xi^{0\nu}_2 + 
\varphi^{\nu}_2  \, \chi^{\nu}_2  \, \xi^{0\nu}_1) + c_{\nu}  \, 
(\xi^{\nu}_2  \, \xi^{0\nu}_1 + \xi^{\nu}_1  \, 
\xi^{0\nu}_2)  \, \sigma^{\nu} \\\nonumber
&+& d_{\nu}  \, (\varphi^{\nu}_1  \, \xi^{\nu}_2  \, \chi^{0\nu}_1 
+ \varphi^{\nu}_2  \, \xi^{\nu}_1  \, \chi^{0\nu}_2) + f_{\nu}  \, 
((\varphi^{\nu}_1)^2  \, \chi^{0\nu}_2 + (\varphi^{\nu}_2)^2  \, 
\chi^{0\nu}_1) + g_{\nu}  \, (\chi^{\nu}_2  \, \chi^{0\nu}_1 + 
\chi^{\nu}_1  \, \chi^{0\nu}_2)  \, \sigma^{\nu} \, ,
\end{eqnarray}
we obtain a minimum of the potential by setting 
the F-terms of the driving fields to zero:
\begin{eqnarray}\nonumber
\frac{\partial w_{f,\nu}}{\partial \psi^{0\nu}} &=& a_{\nu}  \, 
(\chi^{\nu}_1  \, \xi^{\nu}_1 - \chi^{\nu}_2  \, \xi^{\nu}_2) = 0 
\, ,\\ \nonumber
\frac{\partial w_{f,\nu}}{\partial \chi_1^{0\nu}} &=& d_{\nu}  \, 
\varphi^{\nu}_1  \, \xi^{\nu}_2  + f_{\nu}   \, 
(\varphi^{\nu}_2)^2 + g_{\nu}  \, \chi^{\nu}_2  \, 
\sigma^{\nu}  = 0 \, ,\\ \nonumber
\frac{\partial w_{f,\nu}}{\partial \chi_2^{0\nu}} &=& d_{\nu}  \, 
\varphi^{\nu}_2  \, \xi^{\nu}_1  + f_{\nu}   \, 
(\varphi^{\nu}_1)^2 + g_{\nu}  \, \chi^{\nu}_1  \, \sigma^{\nu}   
= 0 \, ,\\ \nonumber
\frac{\partial w_{f,\nu}}{\partial \xi_1^{0\nu}} &=& b_{\nu}  \, 
\varphi^{\nu}_2  \, \chi^{\nu}_2  + c_{\nu}  \, \xi^{\nu}_2  \, 
\sigma^{\nu} = 0 \, ,
\\ \nonumber
\frac{\partial w_{f,\nu}}{\partial \xi_2^{0\nu}} &=& b_{\nu}  \, 
\varphi^{\nu}_1  \, \chi^{\nu}_1  + c_{\nu}  \, \xi^{\nu}_1  \, 
\sigma^{\nu}  = 0 \, .
\end{eqnarray}
As advocated above, these two sets of equations are uniquely solved 
by the VEV configurations given in 
Eqs.~(\ref{eq:VEVs_e}) and (\ref{eq:VEVs_nu}), where we have 
set a possible relative phase in the doublet of VEVs of the 
flavons in the charged lepton sector to zero. 
This can be done without loss of generality, as only the 
phase difference between the two sectors is phenomenologically 
relevant. We have also assumed that none of the parameters 
in the superpotential vanish. For the charged lepton sector, the flavon VEVs 
$w_e$ and $x_e$ are free parameters (which we take to be non-zero), while 
\begin{equation}
\label{eq:value_we_ze}
v_e  =  e^{\frac{4 \pi i k}{5}} \, 
\frac{c_e \, d_e \, x_e}{b_e \, f_e} ~, ~~ 
z_e  =  e^{\frac{8 \pi i k}{5}} \, \frac{d_e \, x_e}{f_e} \, .
\end{equation}
Similarly $v_{\nu}$ and $x_{\nu}$ are 
free parameters (again taken to be non-vanishing) and 
\begin{equation}
\label{eq:value_wnu_znu}
w_{\nu}  =  -\frac{c_{\nu} \, f_{\nu} \, x_{\nu} \, v_{\nu}^2}
{c_{\nu} \, g_{\nu} \, x_{\nu}^2 - b_{\nu} \, d_{\nu} \, 
v_{\nu}^2} \; , \;\; 
z_{\nu}  =  \frac{b_{\nu} \, f_{\nu} \, v_{\nu}^3}{c_{\nu} \, 
g_{\nu} \, x_{\nu}^2 - b_{\nu} \, d_{\nu} \, v_{\nu}^2} \, .
\end{equation}
The driving fields themselves are only allowed 
vanishing VEVs, as can be inferred from considering the 
F-terms of the flavons. 
Note, that since we can not 
make the cutoff scale $\Lambda$ arbitrarily 
large, we need to take into account NLO corrections to 
both the Yukawa and flavon superpotentials. We also should be careful
in what regards potentially dangerous flavor changing neutral currents
induced by the flavons. All this could be taken into account by
carefully studying the mass spectrum of the scalars. 
Given the sizable number of
fields this is a formidable task, but fortunately it suffices to 
make some general estimates, which agree well 
quantitatively with a lengthy explicit calculation in a 
similar model \cite{corr}: the $\tau$ lepton 
mass, see Eq.~(\ref{eq:mleps}), 
is of order $\langle f \rangle  \, v /\Lambda$, 
where $\langle f \rangle$ is a flavon vev, $v$ the Higgs 
vev ($\simeq 10^2$ GeV) and $\Lambda$ the cutoff scale. 
The neutrino mass, see Eq.~(\ref{eq:mnus}), 
is of order $\langle f \rangle  \, v^2 /\Lambda^2$. 
With the charged lepton $\tau$ mass $\simeq$ GeV and the neutrino mass 
$\simeq 0.1$ eV it follows $\Lambda \simeq  10^{12}$ GeV and 
$\langle f \rangle \simeq 10^{10}$ GeV. Now we can
estimate that the flavon mass is also 
of order of $\langle f \rangle$. NLO corrections to the potential, and
therefore to the neutrino and charged lepton mass matrices, are of
order $\langle f \rangle/\Lambda \simeq 10^{-2}$ and therefore 
under control. 
Any potentially dangerous flavor changing 
neutral currents are also suppressed by the heavy mass scale 
$\langle f \rangle$.

\end{appendix}

\end{document}